\documentclass[conference]{IEEEtran}
\IEEEoverridecommandlockouts
\IEEEpubid{\makebox[\columnwidth]{ 979-8-3503-5067-8/24/\$31.00~\copyright2024 IEEE \hfill}
\hspace{\columnsep}\makebox[\columnwidth]{ }}
\usepackage{cite}
\usepackage[utf8]{inputenc}
\usepackage[T1]{fontenc}
\usepackage{amsmath,amssymb,amsfonts}
\usepackage{algorithmic}
\usepackage{graphicx}
\usepackage{textcomp}
\usepackage{xcolor}
\usepackage{svg}
\usepackage{tabularx}
\def\BibTeX{{\rm B\kern-.05em{\sc i\kern-.025em b}\kern-.08em
    T\kern-.1667em\lower.7ex\hbox{E}\kern-.125emX}}
\begin{document}

\title{Investigating Trading Mechanisms as a Driver for User Experience in Racing Games}

\author{\IEEEauthorblockN{Georg Arbesser-Rastburg}
\IEEEauthorblockA{
\textit{Graz University of Technology}\\
Graz, Austria \\
georg.arbesser-rastburg@tugraz.at}
\and
\IEEEauthorblockN{Thomas Olip}
\IEEEauthorblockA{
\textit{Graz University of Technology}\\
Graz, Austria \\
thomas.olip@gmail.com
}
\and
\IEEEauthorblockN{Johanna Pirker}
\IEEEauthorblockA{
\textit{Ludwig-Maximilians-Universität München}\\
Munich, Germany \\
jpirker@iicm.edu}
}

\maketitle
\IEEEpubidadjcol

\begin{abstract}
The exchange of digital goods has become a significant aspect of the global economy, with digital products offering inexpensive reproduction and distribution. In-game objects, a type of digital currency, have emerged as tradable commodities within gaming ecosystems. Despite extensive research on various aspects of digital goods, little attention has been given to the impact of in-game trading mechanisms on user experience. This paper presents a study aimed at evaluating the influence of trading systems on user experience in a racing game context. We developed a simple racing game featuring an in-game market for buying and selling car variants and conducted an A/B study comparing user experiences between groups utilizing the trading system and those unlocking cars through race completion. Our findings suggest that while the trading system did not significantly alter the overall user experience, further exploration of diverse trading approaches may offer insights into their impact on user engagement.
\end{abstract}


\section{Introduction} \label{sec:intro}
The exchange of digital goods has impacted the global economy. Digital goods are traded between producers/sellers and consumers in digital formats, making them inexpensive to reproduce and distribute \cite{Bradley2012TowardBusiness}.

One type of digital goods is tradable in-game objects, which can be seen as digital currency \cite{Adams2012GameDesign}. These in-game objects can be used to purchase goods and services within the game, and popular games like the Pokémon series\footnote{www.pokemon.com/us/pokemon-video-games} or Massively Multiplayer Online Role-Playing Games (MMORPGs) like World of Warcraft\footnote{worldofwarcraft.blizzard.com} have incorporated trading systems to facilitate this process. For instance, Forza Horizon 5\footnote{www.forza.net/horizon} has an auction house where players can offer their vehicles for sale.

The field of trading in digital games has been studied in various aspects. Researchers have explored how markets outside the actual games can emerge and affect the game itself \cite{Guo2007WhyMoney}. Furthermore, in addition to trading virtual goods, current games allow players to make in-game purchases using real money. Researchers have investigated various aspects of this phenomenon, including its impact on game balance and player behavior \cite{Urschel2011UnderstandingMMORPGs}. Especially the effects of loot boxes in games have been researched in the past years, for instance, the relationship between loot boxes and gambling \cite{Li2019TheGambling}.

However, despite the extensive research in the digital goods industry, there has yet to be research on the effect of in-game trading mechanisms on the user experience. Therefore, this paper aims to take a first step toward evaluating whether integrating a trading system significantly influences the user experience. To achieve this, we implemented a simple racing game that allows players to buy and sell car variants in an in-game market. By analyzing the user feedback, we aim to provide insights into the impact of in-game trading mechanisms on the user experience.

We describe the racing game in Section \ref{sec:game}. Using this game, we ran an A/B study where one group used the in-game market, whereas the other group unlocked new cars by simply completing races. We describe the study in Section \ref{sec:evaluation} and discuss our findings, the study's limitations, and potential future research in Section \ref{sec:discussion}. More information about the work can be found in \cite{Olip2024ImplementationGame}.
\section{Racing Game Implementation} \label{sec:game}
To determine how trading mechanisms affect user experience in racing games, we implemented a simple racing game in Unity\footnote{www.unity.com}. In addition, we used the Vehicle Physics Pro script\footnote{evp.vehiclephysics.com} as a basis for the vehicle physics implementation. The goal for the players is to reach speeds as high as possible on a German highway without crashing into other vehicles or the environment.

Currently, the game contains one highly detailed sports car (called \textit{Teron Kanaani}) that is available for racing. While only a single vehicle is available, it comes in different colors that, for now, serve as a substitute for different cars.

\subsection{Game Menus}
When players start the game, they are taken to the main menu. From here, they can access several submenus:

\begin{itemize}
    \item Collection
    \item Garage
    \item Market
    \item Profile
    \item Events
\end{itemize}

\begin{figure}
    \centering
    \includegraphics[width = 0.485\textwidth]{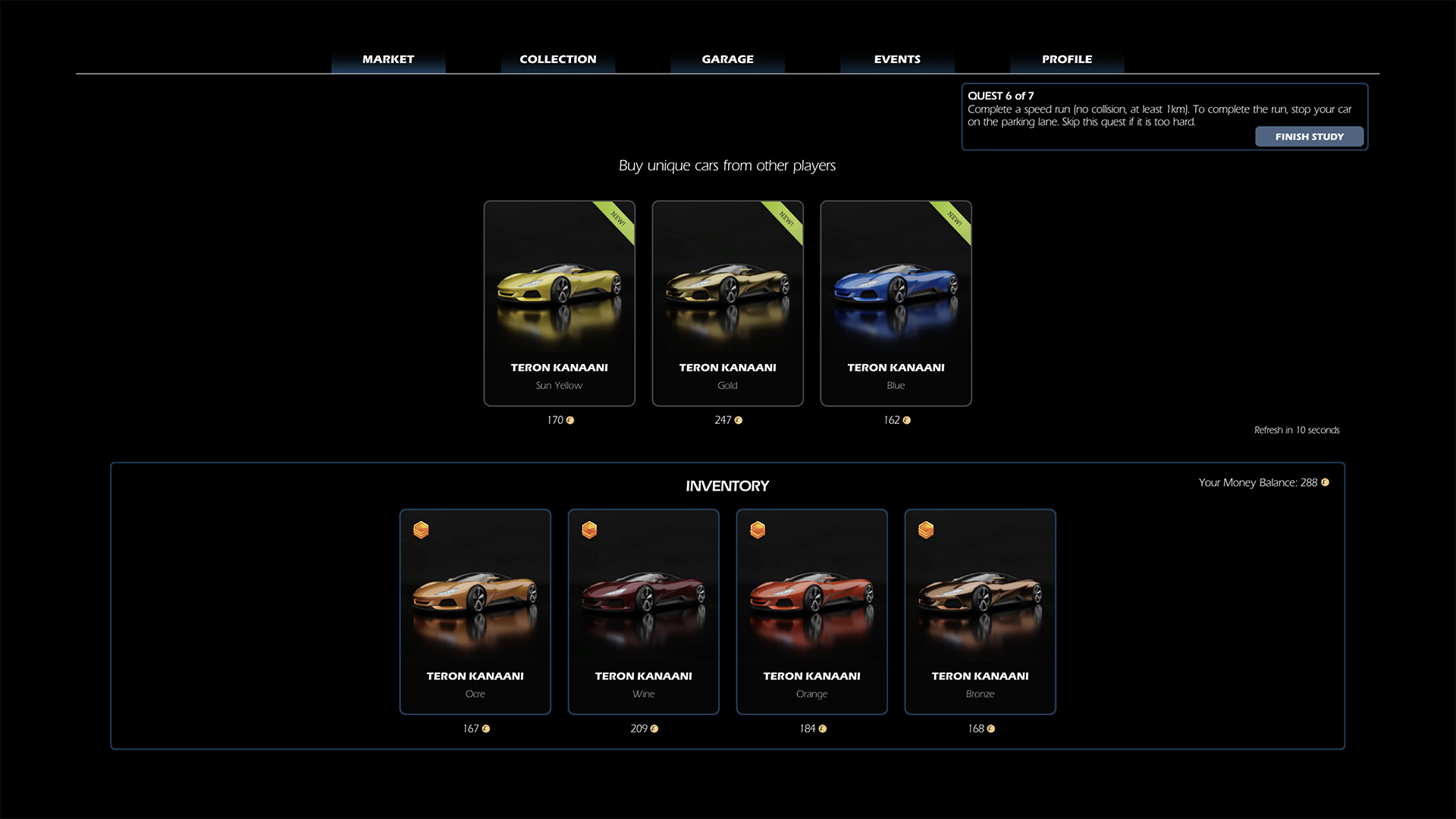}
    \caption{Screenshot depicting the in-game market. Players can sell their vehicles in the bottom half and buy new ones in the top half.}
    \label{fig:market}
\end{figure}

The \textit{collection} displays the different available vehicle variants. The players can view unseen, encountered, and owned vehicles, which are visualized differently to give them an overview of their progress. The \textit{garage} shows the player's currently selected vehicle in a simple island environment. The \textit{market} allows players to sell and buy cars. At the moment, players cannot yet trade with each other and can only sell their cars for a fixed price and buy vehicles from a random selection in a simulated market (see Figure \ref{fig:market}). The \textit{profile} menu shows different player statistics, such as the number of owned cars, total distance traveled, and the number of completed races. It also displays various achievements, which come in three levels (bronze, silver, and gold) and reward the player in nine categories (e.g., total traveled distance, number of sold cars, or number of collected vehicles). Finally, the \textit{events} menu allows players to start new races. However, currently, only highway races are available.

\subsection{Highway Races}
The highway is procedurally generated on runtime by concatenating blocks containing the road and decorations like concrete road barriers or sound barriers. The individual segments can be spawned in a predefined order to facilitate the transition between blocks with different numbers of lanes. Cars populate the highway as obstacles for the players.

These cars are dynamically spawned on the road out of the player's sight to optimize performance. Players must overtake without colliding with them or the environment. One special type of car is police cars that also drive on the highway. If the player collides with a police car, its blue emergency lights are turned on, and roadblocks are created in front of the player to stop the player's vehicle. However, any collision, whether with a regular car, a police car, or the environment, automatically invalidates the current race, and the player will not receive any rewards for the race.

\begin{figure}
    \centering
    \includegraphics[width = 0.485\textwidth]{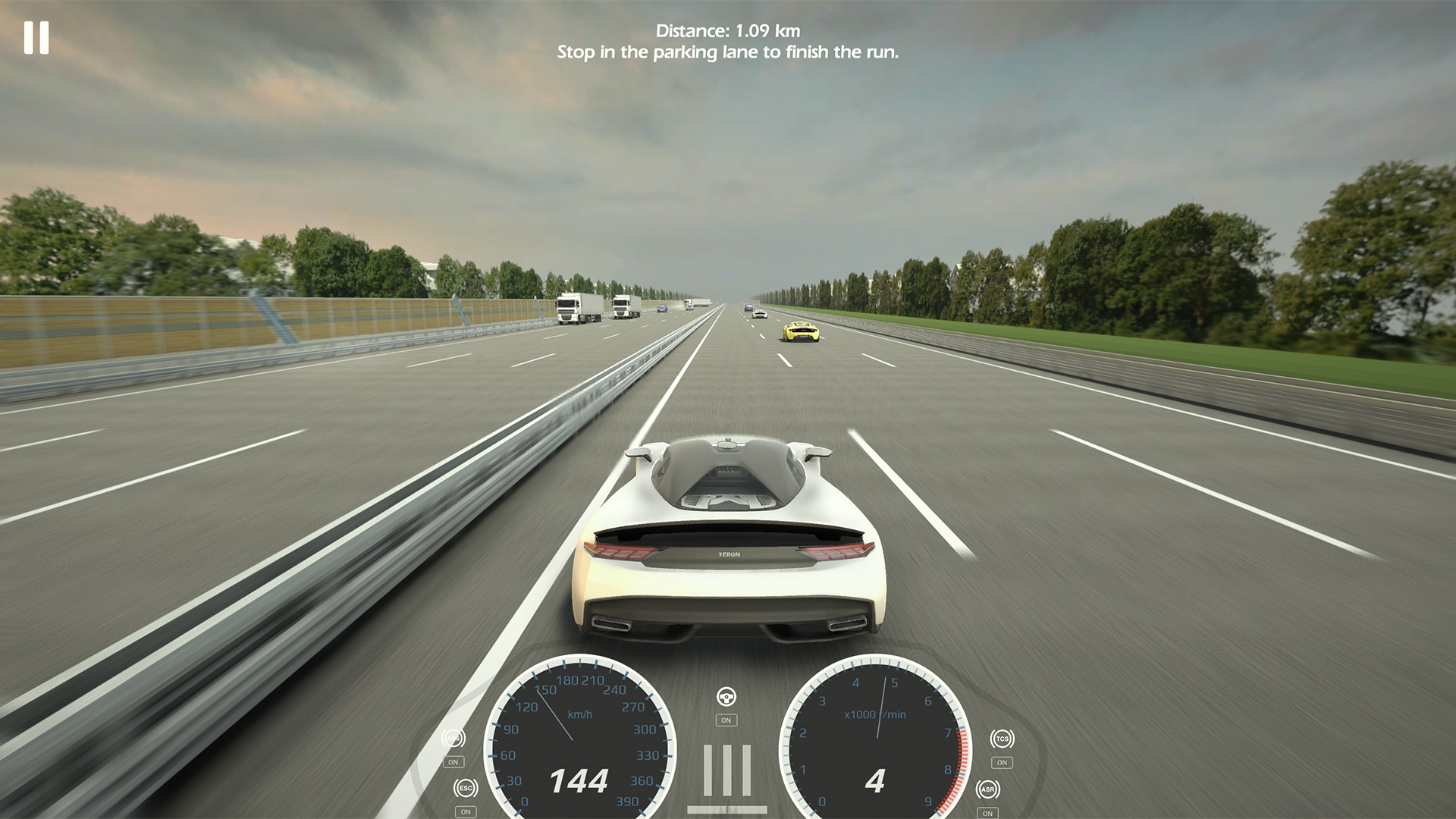}
    \caption{Screenshot taken from a race on the highway with a HUD on the bottom and the traveled distance on top of the screen.}
    \label{fig:race}
\end{figure}

To complete a race, the player has to leave the highway and bring the car to a standstill. Therefore, the player has to choose whether to keep driving and risk crashing while trying to beat their current maximum speed or leave the highway to finish the race. A screenshot from a race can be seen in Figure \ref{fig:race}.

\section{Evaluation} \label{sec:evaluation}
We conducted an A/B study to assess how trading affects user experience in racing games. Participants were asked to complete a pre- and post-questionnaire, and in-game data was collected.

\subsection{Methodology}
Two modes of obtaining new car variants were used in the study:

\begin{itemize}
    \item Mode A: Players receive a random vehicle after completing a race.
    \item Mode B: Players must use in-game currency to purchase new car variants or trade in one of their existing ones.
\end{itemize}

In Mode A, the in-game store is disabled, while in Mode B, players are required to use it. In Mode B, players can only own up to ten cars, so they must trade in an existing one to obtain a new vehicle at some point. Additionally, players in Mode A receive an initial car, while players in Mode B receive an initial budget for purchasing cars.

\subsubsection{Quest System}
To guide users through the study, we implemented a simple quest system. The current quest number and the quest's description were displayed in the top right corner of the user interface. Table \ref{table-quests} provides a complete list of all quests.

\begin{table}[!ht]
\caption{\label{table-quests} Quests during the study per group.}
\begin{center}
\def\arraystretch{1.5}%
\begin{tabularx}{0.5\textwidth} { 
   >{\centering\arraybackslash} p{0.06\textwidth}
   >{\centering\arraybackslash} p{0.03\textwidth}
   >{\raggedright\arraybackslash}X }
  \hline
Quest Number & Test Group & Quest Description\\
 \hline
 1. & A & Select your new car in the 'Collection' tab.\\
 
 1. & B & Visit the 'Market', buy a car, and select it in the 'Collection' tab.\\
 
 2. & A/B & Start a speed run in the 'Garage' tab. Drive at least 1 km.\\
 
 3. & A & Change your car in the 'Collection' tab.\\
 
 3. & B & Visit the 'Market' and buy another car.\\
 
 4. & A/B & Start a speed run and drive at least 2 km. Try not to crash!\\
 
 5. & A & Change your car again in the 'Collection' tab.\\

 5. & B & Visit the 'Market' and buy another car.\\
 
 6. & A/B & Complete a speed run (no collision, at least 1km). To complete the run, stop your car in the parking lane. Skip this quest if it is too hard.\\
 
 7. & A/B & Feel free to play some more races. When you are done, press 'Finish'.\\
  \hline
\end{tabularx}
\newline
\end{center}
\end{table}

\subsubsection{Procedure}
A link to the game was distributed online in Discord groups or directly provided to interested persons. Using the link, users could download the game, which supports Windows, macOS, and Linux. After starting the game, users had to complete the pre-questionnaire and were randomly assigned to group A or B. Afterward, they had to complete their quests, and finally, the users were asked to fill out the post-questionnaire.

\subsubsection{In-Game Data Collection}
During the game sessions, we collected in-game data to evaluate the number of cars obtained by the players and the number of interactions with different elements in the game. The data was automatically sent to a database for later evaluation. A complete list of the collected in-game data is provided in Table \ref{table-data-collection}.

\begin{table}[!ht]
\caption{\label{table-data-collection}List of in-game data collected during the study.}
\begin{center}
\def\arraystretch{1.5}%
\begin{tabularx}{0.5\textwidth} { 
   >{\centering\arraybackslash}l 
   >{\raggedright\arraybackslash}X }
  \hline
Name & Description\\
 \hline
 Play Time & Total time spent in the program, measured in seconds.\\

 Distance & Combined driven distance with the cars, measured in meters.\\
 
 Races & Total driven races, including aborted races.\\
 
 Finished Runs & Total finished races without crashes or disqualification.\\
 
 Cars Collected & Total cars obtained through winning or trading.\\
 
 Market Visits & Clicks on the market tab in the user interface.\\
 
 Collection Visits & Clicks on the collection tab in the user interface.\\
 
 Garage Visits & Clicks on the garage tab in the user interface.\\
 
 Event Visits & Clicks on the events tab in the user interface.\\
 
 Profile Visits & Clicks on the profile tab in the user interface.\\
 
 Finished Quests & Number of quests the user completed.\\
  \hline
\end{tabularx}
\newline
\end{center}
\end{table}

\subsubsection{Questionnaires}
In addition to the in-game data collected, we asked the study participants to fill out a pre- and a post-questionnaire. Both were integrated directly into the game. The pre-questionnaire only contained questions regarding the players' age and sex. The post-questionnaire consisted of the System Usability Scale (SUS) \cite{Brooke1996SUS:Scale} and the Game Experience Questionnaire (GEQ) \cite{Ijsselsteijn2013TheQuestionnaire}. We used the shortened In-Game module of the GEQ to keep the questionnaire brief.

\subsubsection{Participants}
39 participants, eight female and 31 male, took part in the study. The participants' ages ranged from 14 to 41, with an average of 25.0 and a standard deviation of 6.1.

\subsection{Results}
Out of the 39 participants, 30 filled out the questionnaires.

\subsubsection{System Usability Scale}
We excluded three participants from the SUS analysis due to inconsistent answers. The SUS comprises alternating positively and negatively framed questions. If participants choose the same answer for all questions, this could be due to either not understanding the questions correctly or a lack of attention. Two participants answered all questions the same, while another chose almost the same responses for all questions.

Table \ref{table-sus-results} shows the SUS results per group and the result of a two-sample t-test. While Group A has a lower mean (69.29) than Group B (80.54), the higher standard deviation in Group A (22.69) compared to Group B (7.73) leads to inconclusive results, which is confirmed by the t-test.

\begin{table}[ht]
\caption{\label{table-sus-results}System Usability Score results, including the mean and standard deviation per group, and the statistical significance with $\alpha$ = 0.05.}
\begin{center}
\def\arraystretch{1.5}%
\begin{tabularx}{0.5\textwidth} { 
   >{\raggedright\arraybackslash} p{0.15\textwidth}
   >{\raggedright\arraybackslash} X
   >{\raggedright\arraybackslash} X
   >{\raggedright\arraybackslash} X
   >{\raggedright\arraybackslash} X
   >{\raggedright\arraybackslash} c
   >{\raggedright\arraybackslash} c
   >{\raggedright\arraybackslash} c }
\hline
 & \multicolumn{2}{c}{Group A} & \multicolumn{2}{c}{Group B}  & \textit{t}(26) & \textit{p}\\
 \cline{2-5}
 & \textit{M} & \textit{SD} & \textit{M} & \textit{SD}\\
 \hline
 System Usability Score & 69.29 & 22.69 & 80.54 & 7.73 & 1.76 & .090\\
 \hline
\end{tabularx}
\end{center}
\end{table}

\subsubsection{Game Experience Questionnaire}
The 14 questions that the GEQ comprises can be grouped into seven different categories, which are shown together with their results in Table \ref{table-geq}. In general, negative affect received a low rating, while positive affect, challenge, and competence were rated relatively high. All other categories are in the midfield.  Due to the large standard deviations and similar mean values, the results are inconclusive, however, and no differences between the groups can be observed.

\begin{table}[ht]

\caption{\label{table-geq}Game Experience Questionnaire results, including the mean and standard deviation per Group, and the statistical significance with $\alpha$ = 0.05.}
\begin{center}
\def\arraystretch{1.5}%
\begin{tabularx}{0.5\textwidth} { 
   >{\raggedright\arraybackslash} p{0.22\textwidth}
   >{\raggedright\arraybackslash} X
   >{\raggedright\arraybackslash} X
   >{\raggedright\arraybackslash} X
   >{\raggedright\arraybackslash} X
   >{\raggedright\arraybackslash} c
   >{\raggedright\arraybackslash} c
   >{\centering\arraybackslash} c}

 \hline
 & \multicolumn{2}{c}{Group A} & \multicolumn{2}{c}{Group B} & \textit{t}(29) & \textit{p} \\
\cline{2-5}
 & \textit{M} & \textit{SD} & \textit{M} & \textit{SD}\\
 \hline
Competence & 2.46 & 0.93 & 2.59 & 1.00 & -0.35 & .726\\

 Sensory and Imaginative Immersion & 1.96 & 0.84 & 2.09 & 1.15 & -0.33 & .740\\

 Flow & 2.04 & 1.03 & 2.18 & 1.29 & -0.33 & .742\\

 Tension & 1.54 & 0.91 & 1.85 & 1.30 & -0.77 & .446\\

Challenge & 2.82 & 0.67 & 2.38 & 1.07 & 1.34 & .192\\

Negative affect & 0.96 & 0.77 & 1.71 & 1.46 & -1.71 & .098\\

Positive affect & 2.46 & 0.87 & 2.41 & 1.20 & 0.14 & .892\\
 \hline
\end{tabularx}
\newline
\end{center}

\end{table}

\subsubsection{In-Game Data}
Finally, Table \ref{table-player-stats} shows the collected in-game data. Both the number of market visits and cars sold is 0 for Group A, as this group had no access to the in-game store. Aside from the number of cars sold and the number of market visits, one statistically significant difference between the two groups can be observed: the number of collected cars.

\begin{table}[ht]

\caption{\label{table-player-stats}Collected in-game data including mean and standard deviation per group, and the statistical significance with $\alpha$ = 0.05.}
\begin{center}
\def\arraystretch{1.5}%
\begin{tabularx}{0.5\textwidth} { 
   >{\centering\arraybackslash} l
   >{\raggedright\arraybackslash} X
   >{\raggedright\arraybackslash} l
   >{\raggedright\arraybackslash} X
   >{\raggedright\arraybackslash} l
   >{\raggedright\arraybackslash} c
   >{\raggedright\arraybackslash} c
   >{\centering\arraybackslash} c }
\hline
 & \multicolumn{2}{c}{Group A} & \multicolumn{2}{c}{Group B}  & \textit{t}(33) & \textit{p}\\
 \cline{2-5}
 & \textit{M} & \textit{SD} & \textit{M} & \textit{SD}\\
 \hline
Market Visits & 0.00 & 0.00 & 11.61 & 10.98 & -4.36 & $<$ .001\\
 
 Collection Visits & 17.65 & 10.94 & 17.17 & 11.80 & 0.13 & .901\\

 Garage Visits & 28.00 & 18.14 & 18.00 & 10.75 & 2.00 & .054\\
 
 Event Visits & 5.94 & 8.30 & 2.61 & 3.97 & 1.53 & .136\\

 Profile Visits & 2.88 & 3.66 & 1.11 & 1.23 & 1.94 & .060\\

 Distance (km) & 55.71 & 60.99 & 26.50 & 25.39 & 1.87 & .070 \\

 Play Time (min) & 43.21 & 40.22 & 23.81 & 13.42 & 1.93 & .061\\

 Races & 27.53 & 26.43 & 14.44 & 10.01 & 1.96 & .058\\

 Finished Runs & 2.24 & 2.02 & 1.50 & 1.42 & 1.25 & .219\\

 Cars Collected & 11.53 & 6.35 & 4.94 & 3.42 & 3.85 & .001\\

 Cars Sold & 0 & 0 & 1.61 & 3.09 & -2.15 & .039\\
 \hline
\end{tabularx}
\newline
\end{center}

\end{table}

\section{Discussion} \label{sec:discussion}
The general outcome of the study is positive, with both SUS and GEQ showing positive overall results. SUS indicates average to above-average system usability \cite{Lewis2018ItemScale}. Another interesting aspect is the time spent by the players. While the study itself could be completed in 10 to 15 minutes, many players spent much more time in the game and completed more races than was required. This suggests that the game was engaging and enjoyable for the players. Interestingly, the players did not exhaust the limited number of garage slots for Group B; still, they sold more cars than required by the quests.  This indicates a high level of interest in the game's mechanics and a willingness to explore the game beyond the required tasks.

The study also revealed no statistically significant differences in player experience or system usability between the two groups, except for the number of collected cars, which was higher in Group A. This finding suggests that the game's mechanics were equally enjoyable and usable for both groups. 

\subsection{Limitations \& Future Work}
The study presented in this paper has certain limitations that need to be acknowledged. First, the sample size of the study participants is relatively small, and a larger number of participants would increase the statistical power of the results. 

Second, the study participants could only trade with a simulated market. Introducing a social element, such as allowing participants to trade with each other directly, might lead to different outcomes and provide a more realistic representation of trading behavior. 

Third, players could only acquire the same car in different colors, which may not have motivated them enough to trade cars. Offering different cars with varying stats for acceleration or top speed could increase the motivation to trade. Additionally, integrating car upgrades could make the trading process more interesting and engaging for the players. 

Finally, the rarity of certain cars could play a role in incentivizing players to trade more common cars for rare ones. Being able to collect rare cars might be an incentive for players to trade more frequently and make the trading process more exciting. Overall, these limitations need to be considered when interpreting the study results and should be addressed in future research to provide a more comprehensive understanding of trading behavior in racing games.

\subsection{Conclusion}
In this study, we developed a racing game with a simple trading system to investigate whether trading affects user experience. In the game, players can buy new car variants in an in-game store and race on a highway. During these races, players try to beat their own top speeds without colliding with other vehicles or the environment. To finish the race, players have to bring their cars to a standstill. We tested this system in an A/B study where one group unlocked a new car variant after each race, and the other group had to use an in-game store to buy and sell cars. The results indicate a good overall user experience, but no statistically significant difference between the groups could be observed. This could be due to the fact that players could only buy different variants of one car without any performance differences, and they could only trade with a simulated market. Based on our findings, we conclude that a simple trading system like this does not have a significant impact on user experience. Further research could explore the influence of different trading approaches on user experience.

\bibliographystyle{IEEEtran}
\bibliography{references_manual}

\begin{thebibliography}{1}
\providecommand{\url}[1]{#1}
\csname url@samestyle\endcsname
\providecommand{\newblock}{\relax}
\providecommand{\bibinfo}[2]{#2}
\providecommand{\BIBentrySTDinterwordspacing}{\spaceskip=0pt\relax}
\providecommand{\BIBentryALTinterwordstretchfactor}{4}
\providecommand{\BIBentryALTinterwordspacing}{\spaceskip=\fontdimen2\font plus
\BIBentryALTinterwordstretchfactor\fontdimen3\font minus \fontdimen4\font\relax}
\providecommand{\BIBforeignlanguage}[2]{{%
\expandafter\ifx\csname l@#1\endcsname\relax
\typeout{** WARNING: IEEEtran.bst: No hyphenation pattern has been}%
\typeout{** loaded for the language `#1'. Using the pattern for}%
\typeout{** the default language instead.}%
\else
\language=\csname l@#1\endcsname
\fi
#2}}
\providecommand{\BIBdecl}{\relax}
\BIBdecl

\bibitem{Bradley2012TowardBusiness}
S.~Bradley, C.~Kim, J.~Kim, and I.~Lee, ``{Toward an evolution strategy for the digital goods business},'' \emph{Management Decision}, vol.~50, no.~2, pp. 234--252, 3 2012.

\bibitem{Adams2012GameDesign}
E.~E.~W. Adams and J.~Dormans, \emph{{Game mechanics : advanced game design}}.\hskip 1em plus 0.5em minus 0.4em\relax New Riders, 2012.

\bibitem{Guo2007WhyMoney}
\BIBentryALTinterwordspacing
Y.~Guo and S.~Barnes, ``{Why people buy virtual items in virtual worlds with real money},'' \emph{ACM SIGMIS Database: the DATABASE for Advances in Information Systems}, vol.~38, no.~4, pp. 69--76, 10 2007. [Online]. Available: \url{https://dl.acm.org/doi/10.1145/1314234.1314247}
\BIBentrySTDinterwordspacing

\bibitem{Urschel2011UnderstandingMMORPGs}
A.~Urschel, ``{Understanding Real Money Trading in MMORPGs},'' New World Order, Tech. Rep., 2011.

\bibitem{Li2019TheGambling}
W.~Li, D.~Mills, and L.~Nower, ``{The relationship of loot box purchases to problem video gaming and problem gambling},'' \emph{Addictive Behaviors}, vol.~97, pp. 27--34, 10 2019.

\bibitem{Olip2024ImplementationGame}
T.~Olip, ``{Implementation of Trading as a Collecting Mechanic in a Novel Racing Game},'' Master's thesis, Graz University of Technology, Graz, 2024.

\bibitem{Brooke1996SUS:Scale}
J.~Brooke, ``{SUS: A 'Quick and Dirty' Usability Scale},'' in \emph{Usability Evaluation In Industry}.\hskip 1em plus 0.5em minus 0.4em\relax CRC Press, 6 1996, pp. 207--212.

\bibitem{Ijsselsteijn2013TheQuestionnaire}
W.~A. Ijsselsteijn, D.~Kort, and Y.~A. W.~. Poels, ``{The Game Experience Questionnaire},'' Technische Universiteit Eindhoven, Tech. Rep., 1 2013.

\bibitem{Lewis2018ItemScale}
J.~R. Lewis and J.~Sauro, ``{Item Benchmarks for the System Usability Scale},'' \emph{Journal of Usability Studies}, vol.~13, pp. 158--167, 2018.

\end{thebibliography}

\end{document}